# Low temperature structural effects in the (TMTSF)$_2$PF$_6$ and AsF$_6$ Bechgaard salts

P. Foury-Leylekian[a], S. Petit[b], I. Mirebeau[b], G. Andre[b], M. de Souza[c,d], M. Lang[d], E. Ressouche[e], A. Moradpour[a] and J.-P. Pouget[a]

[a] Laboratoire de Physique des Solides, Université Paris-Sud, CNRS UMR 8502, F-91405 Orsay Cedex, France
[b] Laboratoire Léon Brillouin, CEA-CNRS UMR 12, F-91191 Gif-sur-Yvette Cedex, France
[c] Instituto de Geociências e Ciências Exatas - IGCE, Unesp - Univ Estadual Paulista, Departamento de Física, Cx. Postal 178, 13506-900 Rio Claro (SP), Brazil
[d] Physikalisches Institut, Goethe-Universität Frankfurt, SFB/TR 49, D-60438 Frankfurt (M), Germany
[e] SPSMS, UMR-E CEA/UJF-Grenoble 1, INAC, Grenoble, F-38054, France

**Abstract**

We present a detailed low-temperature investigation of the statics and dynamics of the anions and methyl groups in the organic conductors (TMTSF)$_2$PF$_6$ and (TMTSF)$_2$AsF$_6$ (TMTSF : tetramethyl-tetraselenafulvalene). The 4 K neutron scattering structure refinement of the fully deuterated (TMTSF)$_2$PF$_6$-D12 salt allows locating precisely the methyl groups at 4 K. This structure is compared to the one of the fully hydrogenated (TMTSF)$_2$PF$_6$-H12 salt previously determined at the same temperature. Surprisingly it is found that deuteration corresponds to the application of a negative pressure of 5×10$^2$ MPa to the H12 salt. Accurate measurements of the Bragg intensity show anomalous thermal variations at low temperature both in the deuterated PF$_6$ and AsF$_6$ salts. Two different thermal behaviors have been distinguished. Low-Bragg-angle measurements reflect the presence of low-frequency modes at characteristic energies $\theta_E$ = 8.3 K and $\theta_E$ = 6.7 K for the PF$_6$-D12 and AsF$_6$-D12 salts, respectively. These modes correspond to the low-temperature methyl group motion. Large-Bragg-angle measurements evidence an unexpected structural change around 55 K which probably corresponds to the linkage of the anions to the methyl groups via the formation of F…D-CD$_2$ bonds observed in the 4 K structural refinement. Finally we show that the thermal expansion coefficient of (TMTSF)$_2$PF$_6$ is dominated by the librational motion of the PF$_6$ units. We quantitatively analyze the low-temperature variation of the lattice expansion via the contribution of Einstein oscillators, which allows us to determine for the first time the characteristic frequency of the PF$_6$ librations: $\theta_E \approx$ 50 K and $\theta_E$ = 76 K for the PF$_6$-D12 and PF$_6$-H12 salts, respectively.

# I.-Introduction

Discovered more than 30 years ago in the pressurized Bechgaard salt (TMTSF)$_2$PF$_6$ (TMTSF : tetramethyl-tetraselenafulvalene) [1], organic superconductivity, which is now observed in a large panel of organic conductors, remains a major field of research [2-4]. One of the main challenge of present research activities remains the determination of the microscopic origin of the Cooper pairing [5], an issue all the more interesting because of the accumulating evidences in favour of unconventional d-wave symmetry superconductivity at least for the quasi-one-dimensional (quasi-1D) Bechgaard salts [6]. In these latter salts it has been proposed [7] that the d-wave Cooper pairing is achieved through spin-density-wave (SDW) fluctuations which arise because of the close proximity of a SDW ground state. In this scenario, unconventional superconductivity should emerge from a non-Fermi liquid due to the pertinence of SDW correlations [6, 8].

This scenario however requires a detailed description of the metallic state of the Bechgaard salts. Such a description should consider both the various thermal crossovers, affecting the dimensionality of the coherent charge transport, as well as the nature of the incipient density-wave instabilities provoked by the reduced electronic dimensionality in the presence of electron-electron repulsions and/or electron-lattice coupling. In all these aspects (TMTSF)$_2$PF$_6$ already behaves as a complex metal at ambient pressure. At high temperature the coherent charge transport is confined (1D behaviour) to the stack direction *a* where (TMTSF)$_2$PF$_6$ behaves as a Tomonaga-Luttinger liquid [9,10]. In this regime this salt exhibits both 2k$_F$ SDW [11] and bond-order-wave (BOW) [12, 13] 1D instabilities. Upon cooling, the one-particle properties exhibit a 1D to 2D dimensional crossover around 100 K (*T**), the temperature below which the charge transport becomes coherent in the (*a*,*b*) TMTSF layers [14, 15]. For *T* < *T** the warping of the Fermi surface (FS) disables the divergence of the 1D longitudinal 2k$_F$ BOW response [16] which saturates below *T** and then decreases below *T*** ~50K to finally vanish at about *T**** ~ 30 K [13]. In contrast, the 2k$_F$ SDW response, which develops below *T** the inter-stack correlations achieved by the nesting of the warped FS, continues to grow [17]. Then the 2k$_F$ SDW response diverges critically below *T**** [11] to finally stabilize a SDW ground state below $T_{SDW}$ = 12 K.

However, the insulating ground state of (TMTSF)$_2$PF$_6$ is more complex than expected from the stabilization of a pure 2k$_F$ SDW. In particular, 2k$_F$ and 4k$_F$ charge-density-wave (CDW) super-lattice reflections [18-20] have been detected in addition to the 2k$_F$ spin-density-wave modulation. In addition, unexpected features such as the observation of a lattice expansion anomaly at $T_{SDW}$ [21] show that there is a subtle interplay between the magnetism and the lattice degrees of freedom. All these unusual features require a re-examination of the low-temperature structural properties of (TMTSF)$_2$PF$_6$. This is the first purpose of the present study.

The Bechgaard salts (TMTSF)$_2$X are made of zig-zag columns of TMTSF molecules running along the 1D conducting direction *a*. These columns form ($a,b$) layers of donors whose terminal methyl groups form, along the inter-layer direction $c^*$, soft cavities filled by the anions X. The Bechgaard salts crystallize in the triclinic P$\bar{1}$ space group, where in particular each anion is located on an inversion centre of the structure [22]. At ambient conditions, the anions, even centro-symmetric ones such as PF$_6$, as well as the methyl groups, are subject to a considerable thermal motion. Both anion and methyl-group disorders are progressively removed upon cooling. In particular, the refinement of the 20 K and 4 K neutron-scattering structural data of hydrogenated (TMTSF)$_2$PF$_6$ (named PF$_6$-H12 below) [23] show that each anion locks its orientation by establishing two short F-Se contact distances and four F...H-CH$_2$ bonds with four neighbouring TMTSF molecules located nearly in a plane perpendicular to $a$. However, this neutron scattering refinement also shows that probably due to quantum tunnelling, some methyl-groups remain disordered at these low temperatures.

The effects of disorder due to the thermal motions of the anions and the methyl groups were followed by NMR, and in particular by spin-lattice relaxation rate ($T_1^{-1}$) measurements which probe the thermally activated jumps over finite-energy potential barriers. $^1$H $T_1^{-1}$ measurements reveal two methyl group activated processes above 55 K and 74 K [24-27]. These two activated processes are probably due to the presence of different environments of the methyl groups in the structure. In addition an enhancement of relaxation rate was observed upon cooling below 45 K followed by a $T_1^{-1}$ maximum around 20 K [24-26]. It was thus proposed that such low-temperature features could be due to quantum rotational tunnelling which has been evidenced at 1.2 K by $^1$H NMR in slowly cooled (TMTSF)$_2$ClO$_4$ [28]. Measurements of $^{19}$F $T_1^{-1}$ provide evidence of a thermally activated PF$_6$ motion above 140 K [29] (two different activated processes are observed at 135 K and 210 K for the SbF$_6$ anion in (TMTTF)$_2$SbF$_6$ [30] (TMTTF : tetramethyl-tetrathiafulvalene)). Note that these high-temperature activated PF$_6$ motions connect local minima of the anion potential which are not necessarily those of lowest energy in the structure. In particular, a detailed analysis of the anion disorder in (TMTSF)$_2$AsF$_6$ at 125 K (see figure 4 in Ref. [31]) reveals different types of local minima which could be reached by the thermally activated jumps. At lower temperatures, $^{19}$F linewidth measurements in (TMTSF)$_2$PF$_6$ show that PF$_6$ units begin to rotate above 70 K [29] (a similar observationwas made for (TMTTF)$_2$SbF$_6$ [30]). The proximity of this last temperature with the ones, 55K and 74K, of the onset of $^1$H $T_1^{-1}$ activated processes suggests the presence of a cooperative locking of the anions to the methyl protons below these temperatures [26].

NMR relaxation is sensitive to the high-temperature reorientation molecular motions because, as schematically shown in figure 1, $T_1^{-1}$ probes thermally-activated jumps of the PF$_6$ units and the methyl groups over potential barriers height $\Delta$E. However, at low temperature, when these thermally activated molecular motions are blocked, the dynamics is not completely frozen because low-frequency translational or librational oscillations (of energy $\hbar\omega_0$ in figure 1) of the anions and the methyl groups around the minimum of the potential wells remain. In addition to these classical motions, there are evidences that quantum rotations of the methyl groups survive at low temperatures. The second purpose of the present study is to probe these low-energy movements by accurate measurements of the thermal dependence of the intensity of selected Bragg reflections and of their thermodynamics through high-resolution thermal expansion studies. This will allow us to access the low-temperature methyl-group and anion dynamics, a domain which has never been explored in the Bechgaard salts up to now.

In fact, one of the initial motivations of the neutron scattering study was the search of SDW magnetic reflections below $T_{SDW}$ in fully deuterated (TMTSF)$_2$PF$_6$. No such reflections could be detected probably due to the weakness of its itinerant magnetism. However, in the course of these investigations an anomalous thermal dependence of the intensity of several Bragg reflections was discovered. The present paper reports their accurate study together with the determination of the low-temperature structure of (TMTSF)$_2$PF$_6$-D12 (named PF$_6$-D12 below). This study is complemented by high-resolution measurements of the thermal expansion coefficient at low temperature. The paper is organized as follows. The sample preparation and the experimental conditions are given in section II. Section III presents the experimental results. The analysis of the data is given in section IV. The unexpected structural change found around 55 K is separately discussed in section V. General conclusions are drawn in section VI.

## II. Sample preparation and experimental conditions

Accurate Bragg intensity measurements have been performed using single crystal and powder neutron diffraction. For accurate measurements, neutron scattering which does not induce irradiation damages to the Bechgaard salts, in contrast to X-rays [32], is invaluable. The diffraction measurements have been performed on deuterated crystals and powders because of the lower incoherent background and stronger intensity of nuclear scattering of deuterium as compared to hydrogen; the neutron scattering length (in units of $10^{-12}$ cm) which is negative for H, $b_H = -3.74$, becomes positive for D, $b_D = +6.67$.

97.5% deuterated TMTSF was synthesized by our procedure avoiding the use of gaseous H$_2$Se [33], using 3-chloro-2-butanone-$d_7$ prepared according to a recently optimized method [34]. Electro-crystallisations were carried out in dry argon-purged deoxygenated dichloromethane solutions by the usual constant-current procedure. A possible way to increase the size of the resulting single crystals is to decrease the number of the growing seeds; this was accomplished using a lowered electrolyte-salt concentration (2.10$^{-2}$M) along with a very low current density (0.15 µA cm$^2$) for the initial stage of the crystal formations at room temperature. Under these conditions (one to three seeds formed) (TMTSF)$_2$PF$_6$ single crystals from 100 to 350 mg have been obtained.

A PF$_6$-D12 single crystal was characterized by conductivity measurements which clearly show the occurrence of a sharp metal-insulator transition at $T_{SDW}$ = 12.2 K as observed for PF$_6$-H12 [35].

PF$_6$-D12 single crystals were used for the low-temperature neutron scattering refinement of the structure and for the thermal expansion measurements. The thermal evolution studies of the neutron diffraction pattern were performed on ~ 1 g of powder obtained from finely ground single crystals. Powder diffraction patterns were also recorded with PF$_6$-H12 and (TMTSF)$_2$AsF$_6$-D12 (named AsF$_6$-D12 below) powders.

The structure of PF$_6$-D12 was refined from a 4.2 K neutron diffraction data collection (λ = 0.84 Å) performed on the D9 diffractometer of the Institut Laue-Langevin using a displex cryostat with a Joule-Thomson stage. The neutron scattering powder diffraction patterns were mainly recorded on the G6.1 diffractometer (λ = 4.741Å) installed on the cold source at the Orphée-LLB reactor. Additional spectra were also recorded on the 3T1 ($k_F$ = 4.054 Å$^{-1}$) and G4.1 (λ = 2.423 Å) diffractometers installed on the thermal and cold sources, respectively. In these experiments the powder sample was placed in an aluminium or a vanadium can mounted in a $^4$He cryostat enabling to cool down to 1.5 K. The 3T1 experiment was conducted with a 20' Sollers collimator placed on the scattered beam to insure a good resolution and a low background. Slits were installed as close as possible to the cryostat on the incident and scattered beams to keep a low background especially for low scattering angles.

The uniaxial thermal expansion coefficient $\alpha(T)= l^{-1}(\partial l/\partial T)$, where $l$ denotes the sample length, was measured by employing an ultra-high-resolution capacitance dilatometer with a maximum resolution of $\Delta l/l=10^{-10}$, built after [36].

## III. Experimental results

### 1. The 4 K structure of (TMTSF)$_2$PF$_6$-D12 compared to that of (TMTSF)$_2$PF$_6$-H12

The structure of PF$_6$-D12 was refined at 4.2 K in the P$\bar{1}$ space group. The refinement in the non-centrosymmetric P1 space group did not improve the reliability indices (this was also the case for PF$_6$-H12 at 4 K [23]). The neutron data collection parameters and the atomic coordinates of the P$\bar{1}$ structure are respectively given in tables A.1 and A.2 of Annex A. As neutron scattering allows locating precisely the CD$_3$ methyl groups, the full crystal structure, including the refined methyl-group positions, is shown in figure 2.

Table 1 compares the 4 K lattice parameters of PF$_6$-D12 and PF$_6$-H12. It shows that the PF$_6$-D12 lattice parameters are larger than those for PF$_6$-H12. The 4 K lattice parameters of PF$_6$-D12 are comparable to those of PF$_6$-H12 at 120 K [23]. Using the low-temperature compressibility of PF$_6$-H12 [37], one deduces that the increase of volume of PF$_6$-D12 corresponds to the application of a negative pressure of 500 MPa to PF$_6$-H12 at 4 K. This is a much larger effect than the one revealed for κ-(BEDT-TTF)$_2$Cu[N(CN)$_2$]Br where deuteration corresponds to the application of a negative pressure of about 40 MPa to the hydrogenated salt [3].

As for the PF$_6$-H12 salt, the structural refinement of PF$_6$-D12 shows that both CD$_3$ methyl groups and the PF$_6$ anions are ordered at 4 K. The thermal parameters (U(eq) in table A.2) of the P, F, Se and C atoms (nuclei) of the PF$_6$-D12 salt are comparable to those previously determined in PF$_6$-H12 [23]. U(eq) of D is however 50% larger than U(eq) of H, which means that D undergoes larger quantum fluctuations than H. This result is surprising because D has a mass twice larger than that of H. It is tempting to assign this effect to the presence of smaller potential barriers for quantum tunnelling in PF$_6$-D12 than in PF$_6$-H12. In addition, as for PF$_6$-H12 and despite the fact that the PF$_6$ anion is well localized in the 4 K structure, table A.2 shows that U(eq) for F is three times larger than the U(eq) for the central P atom of the anion. This indicates the presence of a residual librational disorder of the anion (which could be induced by the disorder of the methyl groups with which the F is linked via H/D bonds, see below).

The intra-molecular distances of TMTSF and of PF$_6$ are comparable for PF$_6$-D12 and PF$_6$-H12. In addition, the average C-D distance, of 1.091(2)Å in PF$_6$-D12, is close to the average C-H distance, of 1.087Å in PF$_6$-H12. However, the average D-C-D angle, of 107.7(2)° in PF$_6$-D12, is significantly smaller than the average H-C-H angle of 111.5° in PF$_6$-H12.

The inter-molecular distances defined in figure 3 are reported in Table A.3. As discussed in Annex A they vary as expected with the lattice parameter dilatation between PF$_6$-H12 and PF$_6$-D12.

Surprisingly, the behaviour under pressure and upon deuteration differs for the H bonds. While the shortest F…H-CH2 H-bond distances (shown in the bottom panel of figure 2) do not vary under pressure in PF$_6$-H12 [37], significant variations occur upon deuteration of the methyl groups:
- the distance between F(3) and one H/D of the methyl group linked to C(5), F(3)…H(103)/D(103), increases from 2.45 Å to 2.475 Å,
- the distance between F(3) and one H/D of the methyl group linked to C(4), F(3)…H(113)/D(113), increases from 2.46 Å to 2.50 Å,
- the distance between F(1) and the same H/D as above, F(1)…H(113)/D(113), decreases from 2.56 Å to 2.49 Å.

This means that deuteration slightly modifies the geometry of the H/D bond linkage with the anion, and the equalization of the F…D distances shows a re-centring of the anion in the expanded cavity delimited by the CD$_3$ groups.

## 2. Thermal dependence of the Bragg reflection intensity

Figure 4 superimposes two neutron powder diffractograms of PF$_6$-D12 taken at 1.5 K and 60 K. It clearly shows that the intensity of some Bragg reflections changes substantially between these two temperatures. This is unusual because the diffractograms have been taken in a temperature range where thermal parameters (U(eq)) are not expected to vary significantly. More precisely, a substantial positive variation of the intensity of several Bragg reflections ($I_{1.5K}$-$I_{60K}$) can be observed in figure 4. One can also notice a slight shift of the positions of the Bragg reflections at large-Bragg-angles (2θ) due to the weak contraction of the lattice parameters upon cooling.

The thermal dependence of the powder neutron diffractogram has been followed at temperatures below 100 K. Figure 5 gives the thermal variation of the integrated intensity of the isolated (001) reflection located at the smallest reciprocal wave vector **G** (2θ ≈ 20°). Figure 5 shows that the intensity increases upon cooling without any sign of saturation at 1.5 K, the lowest temperature reached. Surprisingly the rate of increase of the intensity is enhanced upon cooling below 30 K. A somewhat similar thermal behaviour is observed in the set of {(0-11), (110), (10-1), (101), (10-1)} superimposed Bragg reflections, labelled (0-11)$_{eq}$ in figure 4 and located at 2θ ≈ 46°, and whose intensity increases by about 25% below ~ 30 K. A different thermal dependence is shown by Bragg reflections located at larger reciprocal wave vectors. For example, figure 6 gives the thermal dependence of the integrated intensity of two different sets of {(210), (-1-21), (1-12)} and {(200), (004)} superimposed Bragg reflections located at 2θ ≈ 84° and ≈ 90°, respectively. They exhibit a quite large increase of intensity of 100% and 35%, respectively, upon cooling. However, in contrast to the (001) reflection, the enhanced rate of increase of the intensity starts at a higher temperature 50-60 K, and the intensity saturates at low temperatures. Several other Bragg reflections also located at large 2θ values exhibit a low-temperature increase of intensity resembling the one shown in figure 6. For example, the intensity of the set of {(-1-21), (1-12)} Bragg reflections increases by about 75% between 60 K and 1.5 K.

In analogy to the data representation for the PF$_6$-D12 salt, figure 7 superimposes the neutron diffraction powder patterns of AsF$_6$-D12 at 1.5 K and 40 K. A comparison with figure 4 shows that the intensity variation occurs for the same Bragg reflections as for PF$_6$-D12. However, the increase in intensity starts at lower temperatures in AsF$_6$-D12. Figure 8 shows the thermal dependence of the (001) reflection of AsF$_6$-D12. The magnitude of the thermal variation is comparable to the one found for the same reflection in PF$_6$-D12 (figure 5), however the increase of the Bragg intensity starts at ~ 20 K in AsF$_6$-D12 compared to ~ 30 K in PF$_6$-D12.

We have also followed the powder neutron diffractograms of (TMTSF)$_2$PF$_6$-H12 as a function of temperature. Unfortunately, the intensity of the Bragg reflections of PF$_6$-H12 is several times weaker than the one of PF$_6$-D12. This feature, probably due to the decrease of the neutron scattering length from the D to the H, together with the increase of the background intensity by one order of magnitude due to the incoherent scattering from the protons, prevents any accurate measurement of the thermal variation of the Bragg intensities.

We have finally followed the temperature dependence of the powder neutron diffractograms of (TMTTF)$_2$PF$_6$-D12. Some Bragg reflections also exhibit substantial variation of their intensity upon cooling. However, their Miller indices are different from those reported above for the anomalous thermal variation in (TMTSF)$_2$PF$_6$-D12. Their thermal variation resembles the ones shown in figure 6 for PF$_6$-D12, however the intensity anomalies in (TMTTF)$_2$PF$_6$-D12 occur at a higher temperature $T_{CO}$ ≈ 84 K which corresponds to lattice changes accompanying the charge-ordering transition [38]. No additional Bragg intensity anomalies can be detected at low temperature. In particular the intensity of the (001) reflection of (TMTTF)$_2$PF$_6$-D12 is still increasing upon cooling but the enhanced rate of increase, occurring below 30 K in (TMTSF)$_2$PF$_6$-D12, is not observed.

### 3. Thermal expansion measurements

Figure 9 gives, for two PF$_6$-D12 samples, the thermal dependence of the linear thermal expansion coefficient, $\alpha_{c^*}$, measured along the inter-layer $c^*$ direction. $\alpha_{c^*}$ has the same magnitude in PF$_6$-D12 and PF$_6$-H12. The thermal dependence of $\alpha_{c^*}$ resembles that previously measured in PF$_6$-H12 [21]. Surprisingly $\alpha_{c^*}$ shows a broad maximum around 150 K and decreases upon further heating in PF$_6$-D12. A similar decrease was also observed upon heating above 100 K in PF$_6$-H12 [21]. An even more pronounced negative contribution to $\alpha_{c^*}$ has been reported for the Fabre salts (TMTTF)$_2$X-H12 with X = PF$_6$, AsF$_6$ and SbF$_6$ above 100 K [39]. A quantitative analysis of the thermal expansion data of the TMTSF salts will be given in section IV.

The inset of figure 9 shows that $\alpha_{c^*}$ exhibits a weak lambda-type anomaly around 12 K, the temperature at which the SDW metal-insulator transition is observed in conductivity measurements performed on the same batch of samples [35]. A thermal expansion anomaly of the same magnitude was previously reported for PF$_6$-H12 at $T_{\text{SDW}}$ [21]. This anomaly bears some resemblance with the specific heat anomaly measured at the SDW transition of PF$_6$-H12 [40].

## IV. Analysis of the data

### 1. Low-temperature structural modifications due to the freezing of anions and methyl groups motion

The data presented in figures 5 and 6 show that the intensity of several Bragg reflections of PF$_6$-D12 anomalously increases upon cooling. It seems, however, that the data exhibit two distinct features. For the Bragg reflections located at low 2θ, the intensity increases below ~ 30 K without apparent saturation (figure 5). For the Bragg reflections located at large 2θ the intensity increases below 50-60 K (noted ~55K below) and saturates at low temperature (figure 6). These distinct effects should have different structural origins for the following reason. Suppose that the thermal dependence of the low-2θ Bragg reflections can be accounted for by a Debye-Waller factor, involving some low-frequency lattice modes (see below). The same modes should induce a drastic thermal reduction of the intensity of the Bragg reflections located at larger 2θ angles because the Debye-Waller factor varies with the square of the reciprocal lattice wave vector G². This is in contradiction with the thermal dependence observed in figure 6. Consequently, another structural effect must be invoked to explain the thermal behaviour of the large-2θ Bragg reflections.

The structural refinement of (TMTSF)$_2$PF$_6$-D12 clearly shows that at 4.2 K D and F atoms exhibit a stronger disorder (i.e. larger U(eq) - see table A.2) than the other atoms. In addition, $^1$H and $^{19}$F NMR studies quoted in the introduction show that the methyl groups and the anions exhibit considerable motions upon heating. In this context, the variation of the Bragg intensities has to be attributed to the progressive disordering of the D and/or the F atoms due to enhanced rotational fluctuations of the methyl groups and/or anions upon heating.

As NMR shows that PF$_6$ stops to rotate below 70 K [29] and that classical methyl-group jumps over their energy barriers ΔE cease below 55 K [24-27], the change of Bragg intensity observed below ~55 K has to be attributed to the ordering of the methyl groups (i.e. to the freezing of their motion) and to the linkage of these methyl groups to the F atoms of the anion by the formation of H-bonds. This lock-in, predicted by the earlier NMR study of ref. [26], is established by the 4 K structural refinements of PF$_6$-D12 (see section III.1) and PF$_6$-H12 [23]. This point will be further discussed in section V. This linkage however does not completely freeze the lattice dynamics as shown by the large U(eq) values found at 4.2K (see table A.2). The non-saturation of the thermal dependence of the reflections observed at low-Bragg-angles shows that low-frequency modes remain active at quite low temperatures. These modes can be associated with the low-frequency classical/quantum motion of the methyl groups which has been also detected by NMR at low temperature [24-26, 28]. Below we estimate the variation of the Bragg intensity due to the progressive freezing of the methyl-group motion.

A quantitative analysis of the freezing of the methyl group motion is presented in Annex B using a simplified model. Below 55K when the classical activated thermal jumps of the methyl groups stop, each methyl group continues to oscillate in the potential well where it is localized (or undergoes a quantum rotation tunnelling between the neighbouring potential wells). In Annex B it is assumed that each CD$_3$ methyl group oscillates rigidly and independently from each other. The D oscillation is characterized by the angular variable φ, whose dynamics is that of a harmonic (Einstein) oscillator of characteristic energy ℏω$_0$ = k$_B$θ$_E$. For harmonic displacements of D, we consider that the probability P(φ) of the φ distribution is a Gaussian of variance σ. Annex B shows that in this case the extra contribution of the D displacement to the Bragg intensity behaves as:

$$\delta I(\mathbf{G}) \propto \exp(-4\sigma^2). \qquad (1)$$

In this expression σ$^2$ is the mean square angular fluctuations which is defined for an Einstein oscillator by expression (C.4) in Annex C:

$$\sigma^2 = <\delta\varphi^2>_T = <\delta\varphi^2>_0 \coth(\theta_E/2T), \qquad (2)$$

where $<\delta\varphi^2>_0$ represents the zero-point fluctuations. The contribution (1), normalized with respect to the $T = 0$ K data, thus behaves as:

$$\exp\{4<\delta\varphi^2>_0[1-\coth(\theta_E/2T)]\}. \qquad (3)$$

Expression (3) fits nicely the thermal dependence of the (001) reflection in PF$_6$-D12 (figure 5) and AsF$_6$-D12 (figure 8), for which the D contributes about 60% of the low temperature Bragg intensity. From the fit one gets $<\delta\varphi^2>_0 = 0.1$ rad$^2$ for both salts and θ$_E$ = 8.3 K for PF$_6$-D12 and θ$_E$ = 6.7 K for AsF$_6$-D12. $<\delta\varphi^2>_0$ leads to a zero-point mean square amplitude fluctuation of D of $<\delta u^2>_0 =$ $r^2<\delta\varphi^2>_0 \sim 0.10$ Å$^2$ (r ≈ 1Å is the radius of the circle containing the three D of the methyl group, see figure B.1). This value consistently amounts to the averaged U(eq) = 0.12 Å$^2$ obtained for D in the 4 K structural refinenent (see table A.2). $\theta_E$ corresponds to a low-frequency mode of about ~ 5 cm$^{-1}$, the frequency of which slightly decreases from PF$_6$-D12 to AsF$_6$-D12. The anions probably influence the D motion by changing either the shape of the potential where the D is located and/or the strength of the H/D-bonding. No such low-frequency modes have been detected in previous optical investigations of (TMTSF)$_2$PF$_6$ (for a recent investigation see [41]).

## 2. Thermal expansion and anion libration

For the analysis of the thermal dependence of the uniaxial thermal expansion coefficient $\alpha_{c*}$ it is better to plot $\alpha_{c*}/T$ as a function of temperature because this quantity is directly connected to the thermal derivative of the entropy $\partial S/\partial T$ (see Annex D). This plot also better illustrates the low-temperature anomalies, resembling specific heat singularities at phase transitions, as found, e.g., at the charge-ordering (CO) and spin-Peierls (SP) transitions for (TMTTF)$_2$PF$_6$-H12 [39]. Moreover from the work in Ref. [39], where for (TMTTF)$_2$PF$_6$ the uniaxial thermal expansion coefficients along the $a$, $b'$ and $c*$ axes where studied as a function of temperature, where $b'$ is perpendicular to the $a$ axis in the ($a$, $b$) plane and $c*$ is perpendicular to the ($a$, $b$) and ($a$, $b'$) planes, it can be seen that the temperature dependence of $\alpha_{c*}(T)/T$ is similar to that of $\beta(T)/T$ where $\beta$ is the volume expansion coefficient. Thus, by analogy, we consider $\alpha_{c*}(T)/T$ to represent the temperature dependence of $\beta(T)/T$ also for PF$_6$-H12 and –D12. Figure 10 compares $\alpha_{c*}/T$ of PF$_6$-D12 (sample IIE119) with the same quantity derived from the data of PF$_6$-H12 [21]. $\alpha_{c*}/T$ of PF$_6$-H12 exhibits a strong thermal dependence which consists of a pronounced maximum at $T_{M1}$ ~ 32 K and a weak shoulder at $T_{M2}$ ~ 43 K. The $\alpha_{c*}/T$ data of PF$_6$-D12 reveal a less intense and broader peak from which one can guess two maxima at $T_{M1}$ ~ 22 K and $T_{M2}$ ~ 36 K plus a shoulder at ~ 60 K. Figure 10 also compares these data with $\alpha_{c*}/T$ obtained for (TMTTF)$_2$PF$_6$-H12 [39] and (TMTTF)$_2$Br-H12 [42]. In particular this figure clearly shows that $\alpha_{c*}/T$ is about one order of magnitude larger in (TMTTF)$_2$PF$_6$-H12 than in (TMTTF)$_2$Br-H12. The only structural difference between these two TMTTF salts being the absence of the anion orientation degree of freedom in the X = Br salt, the data clearly shows that the enhanced thermal expansion of (TMTTF)$_2$PF$_6$-H12 is due to the presence of a large Grűneisen parameter Γ(PF$_6$) (defined in Annex D) associated with the anharmonicity of the anion rotational or librational motions. As shown in figure 10, thermal expansion coefficients of similar magnitude are measured in (TMTTF)$_2$PF$_6$-H12, PF$_6$-H12 and PF$_6$-D12. Thus the anharmonicity of the PF$_6$ anion should also control the thermal expansion of PF$_6$-H12 and PF$_6$-D12.

Below we analyse more quantitatively the thermal dependence of $\alpha_{c*}/T$ in PF$_6$-H12 and PF$_6$-D12. For that purpose we assume that the $\alpha_{c*}/T$ excess compared to the same quantity measured in (TMTTF)$_2$Br-H12 is due to the PF$_6$ anion libration. If one assumes that librations of the PF$_6$ can be accounted for by Einstein oscillators (of energy $k_B\theta_E = \hbar\omega_0$) it is possible to calculate (see Annex

C) their contribution to the entropy and to the specific heat and thus to the thermal expansion via expression (D.4) of Annex D. Figure 11 shows that a single Einstein oscillator with only two adjustable parameters ($\theta_E$ = 76 K plus a pre-factor including the generalized Grüneisen parameter) fits nicely the thermal dependence of $\alpha_{c*}/T$ for $PF_6$-H12 between the SDW transition and 200 K. However, the fit of the thermal dependence of $\alpha_{c*}/T$ for $PF_6$-D12 is not so simple. Firstly, the $T_{M2}$ maximum around 36 K in $\alpha_{c*}/T$ with some indications of a double-peak structure could be an artefact due an enhanced noise level. This extrinsic effect is most likely due to elastic anomalies of oxidized Cu in the body of our dilatometer cell which is contaminating recent thermal expansion measurements in a narrow temperature interval of about 32 – 40 K. Nevertheless a reasonable fit of the thermal dependence of $\alpha_{c*}/T$ for $PF_6$-D12 can be achieved using two Einstein oscillators. The low-frequency Einstein oscillator ($\theta_{EL}$) roughly accounts for the $T_{M1}$ peak position of $\alpha_{c*}/T$ while, at variance to the case of the $PF_6$-H12 sample, a high-frequency Einstein oscillator ($\theta_{EH}$) is required to account for the broadness of the peak and of the reduced rate of decrease of $\alpha_{c*}/T$ at higher temperatures. With this model a least-squares fit to the data taken between 13K and 200K for sample IIE119 (shown in figure 12) yields $\theta_{EL}$ = 46 K and $\theta_{EH}$ = 159 K. The same procedure, applied to $\alpha_{c*}/T$ data for sample IIE111 (not shown) yields $\theta_{EL}$ = 53 K and $\theta_{EH}$ = 214 K. We stress that only $\theta_{EL}$ should be connected with the anion librational motion in its potential well because $\theta_{EH}$ is larger than 140 K, the temperature above which NMR indicates thermally activated reorientations of the $PF_6$ units [29]. The two-Einstein-oscillator fit of the data of the two $PF_6$-D12 samples leads to an anion-libration energy $\theta_{EL} \approx$ 50 K significantly smaller than $\theta_E$ = 76 K found for the single-Einstein-oscillator fit of the $PF_6$-H12 data. This last $\theta_E$ value is close to the temperature of 70 K above which a $^{19}F$ linewidth narrowing due to the anion rotation starts as detected by NMR in $PF_6$-H12 [29].

By associating the shift of $\omega_0$ (= $\theta_E$) between the D12 and H12 salts with the corresponding unit cell volume change quoted in table 1, one estimates, using (D.5), a Grüneisen parameter $\Gamma(PF_6)$ = $-\partial ln\omega_0/\partial lnV$ of about 28 for the $PF_6$ librational mode. This Grüneisen parameter is more than one order of magnitude larger than the one deduced from the frequency shift of the low-frequency TMTSF lattice modes measured by infra-red absorption in $PF_6$-H12 and $PF_6$-D12 [43]. This large $\Gamma(PF_6)$ value consistently confirms that the low-temperature thermal expansion of $(TMTSF)_2PF_6$ is dominated by the librational motion of the $PF_6$ units.

There are 3 librational modes per $PF_6$ of $A_g$ and $B_g$ symmetry which should be Raman active [44]. However, these modes have never been detected by Raman scattering because structural refinements show that the $PF_6$ anion is close to a perfect octahedron of cubic symmetry whose rotation does not change the polarizability tensor. At variance to Raman scattering, it is the large Grüneisen parameter $\Gamma(PF_6)$ which allows us in the present study to probe for the first time the dynamics of the librational modes of the $PF_6$ units via thermal expansion measurements. The frequency of the $PF_6$ librational mode ($\approx$35 $cm^{-1}$ in the D12 and 53 $cm^{-1}$ in the H12) is close to the frequency of the $A_g$ and $B_g$ translational and librational modes of the TMTSF molecule detected by Raman scattering in PF$_6$-H12 [44, 45]. It is also close to the frequency of the A$_u$ and B$_u$ translational and librational TMTSF lattice modes measured by far-infrared absorption in PF$_6$-H12 and PF$_6$-D12 [43]. One thus expects important coupling between all these lattice modes. In particular, translational and librational motions of the TMTSF molecule should modify the geometry of the methyl-group cavity in which the anion rotates and vice-versa.

# V. The structural change around 55 K

In addition to the lattice dynamics of the methyl groups and the anions the effects of which have never been analysed before, another noticeable structural feature shown by our data is the anomalous change around ~55 K of the intensity of several Bragg reflections located at large 2θ (figure 6). In agreement with earlier $^1$H and $^{19}$F NMR studies [24-27, 29] we attribute this change to the ordering or freezing of the methyl group classical motion and to the linkage of these methyl groups to the F atoms of the anion revealed by the formation of H-bonds in the low-temperature structural refinements. The structural change observed at ~55 K could also be detected in earlier measurements with which our structural findings can be interpreted more consistently.

Firstly, the thermal dependence of some lattice parameters exhibits an anomaly around 60 K. For example the triclinic cell angle *γ*, which monotonously decreases upon cooling then further increases below 60 K [23], shows that the linkage of the anions to the terminal methyl groups of the TMTSF leads to a shear deformation of the (*a*, *b*) TMTSF layer. Also lattice perturbations associated to this linkage are revealed, according to the analysis performed in reference [23] of the thermal expansion tensor, by a maximum at ~ 60 K in the principal expansion coefficient $\alpha_{33}$ directed along the long axis of the TMTSF molecule. Note, however, that our uniaxial measurement of the thermal expansion coefficient along the fixed *c** direction does not exhibit a significant anomaly near 60 K (figures 9 and 10).

The net stiffening of the longitudinal sound velocity observed below 55 K [46] shows that the linkage of the PF$_6$ to the methyl groups increases the rigidity of the structure. This increase of lattice rigidity should decrease the electron-phonon coupling between the acoustic like BOW stack deformation and the high temperature 1D 2k$_F$ electronic instability observed in the Bechgaard salts. In particular the 55K lattice hardening nicely explains the drop of intensity of the 2k$_F$ BOW response exhibited by X-ray diffuse scattering data below $T^{**}$ ~ 50 K [12, 13, 18, 19]. Similarly, a drop of the 2k$_F$ BOW diffuse scattering intensity is observed below ~ 50 K in the AsF$_6$ salt [47] which behaves as the PF$_6$ salt.

At the present stage of our investigation there is no clear evidence that the structural modification occurring at ~ 55 K is due to a symmetry breaking. The only symmetry breaking operation expected for a structural transition which keeps the translational triclinic symmetry is the loss of the inversion centres. The loss of inversion centres should lead to a P1 structure similar to the one probably stabilized in the Fabre (TMTTF)$_2$X salts below the CO/ferroelectric transition. In this respect the Bragg intensity variation shown in figure 6 resembles the one observed at the 84 K CO transition in $(TMTTF)_2PF_6$-D12 [38]. However, our structural refinement of $PF_6$-D12 at 4 K, as well as earlier low-temperature refinement performed in $PF_6$-H12 [23] does not provide evidence of a P1 non-centrosymmetric structure. However it should be remarked that the differentiation between P1 and P$\bar{1}$ space groups using the neutron scattering data of $(TMTSF)_2PF_6$ is quite a difficult task. Earlier neutron scattering refinements of the structure of $(TMTTF)_2PF_6$ and $AsF_6$ in its CO/ferroelectric phase were unable to reveal the expected break of inversion symmetry [48]. Finally, the lattice expansion measurements shown in figure 9 do not clearly reveal around ~ 55K either the mean-field-like step anomalies found at the CO transition of $(TMTTF)_2PF_6$ and $AsF_6$ [39] or the lambda-type anomaly found at the CO transition of $(TMTTF)_2SbF_6$ [42].

Although the low-temperature symmetry of $(TMTSF)_2PF_6$ remains debated, we note that the stabilization of a long-range $4k_F$ site CDW or CO consistent with a P1 symmetry seems to be unlikely because the conductivity remains metallic until the SDW transition. However short-range charge disproportionation inducing a local ferroelectric polarization above the SDW transition cannot be excluded. The possibility of a hidden or fluctuating ferroelectricity in the metallic state of the Bechgaard salts has been recently suggested [49] from a re-examination of the optical data. Such a possibility is sustained by the thermal behaviour of the low-frequency dielectric constant of $(TMTSF)_2PF_6$, measured along the interlayer $c^*$ direction, which exhibits a significant increase below 60 K with a rate of increase depending strongly on the measurement frequency [50]. Interestingly the onset of the enhanced charge response coincides with the ~ 55 K structural change depicted in figure 6. In our scenario this means that the H-bond linkage of the anion with the methyl groups probably induces a more polarizable medium revealed by dielectric measurements. The enhanced charge response for electric fields polarizing the methyl group cavity when applied along c* can be explained using a mechanism previously proposed for $(BEDT-TTF)_2X$ salts [51, 52]. A field induced local shift of the anion towards a donor inside the methyl group cavity will increase the H bonding with the donor's methyl groups, which will make the methyl group's hydrogen/deuterium more positive. This in turn will induce a negative σ charge shift leading to an excess of positive π charge in the core of the TMTSF.

## VI. Concluding remarks

The analysis of the thermal dependence of the Bragg intensity and of the lattice expansion has revealed the dynamics related to the $CD_3$ methyl group and $PF_6$ anion librational motions. These low-frequency modes have never been reported before in the literature. Our data complement in the low-temperature range, earlier high-temperature $T_1^{-1}$ NMR spin-lattice relaxation rate measurements which had probed thermally activated jumps over finite potential barriers. As illustrated in figure 1, we have probed the librational motion of the anions and methyl groups localized in their equilibrium potential well when the thermally activated regime was no longer effective. The $PF_6$ units stop their activated jump motion at ~140 K [29], temperature below which they undergo librational motions with a characteristic frequency $\theta_E \sim$ 50 K (D12) -76 K (D12). The methyl groups stop their classical activated jump motion around 55-74 K [24-27]. Below about 55K, the simultaneous localization of the anions and methyl groups in their respective potential wells allows the linkage of the methyl groups with the F atoms of the anion. This accounts for the anomalous intensity variation of several "large-$2\theta$" Bragg reflections around $\sim$ 55 K. This linkage probably occurs through the formation of H/D-bonds which is revealed by our 4 K structural refinement of $PF_6$-D12. However this linkage does not imply that the methyl groups are completely frozen since the NMR relaxation rate was found to further grow below 45 K in order to achieve a maximum of $T_1^{-1}$ around 20 K [24-26]. This behaviour was interpreted as the fingerprint of quantum motions. In the same temperature range, below 30K, a sizeable increase of the intensity of several "small-$2\theta$" Bragg reflections is also observed. In a phenomenological way, this unusual behaviour can be accounted for by a low-frequency oscillator ($\theta_E$ = 8.3 K for $PF_6$-D12 and $\theta_E$ = 6.7 K for $AsF_6$-D12) which mimics the low-temperature methyl-group motion.

The low-temperature effects summarized above concern the methyl group and anion structural degrees of freedom which are located at the periphery of the TMTSF molecule and inside the methyl group cavity, respectively. Although the HOMO conduction path is confined in the core of the TMTSF molecule and between neighbouring molecules via Se-Se short contact distances indicated in figure 3, the methyl and anion lattice degrees of freedom could influence the electronic structure. At first place, the anions could directly contribute to the electronic dispersion of the $\pi$ conduction holes via their Hartree Coulomb potential. However recent DFT calculations of the electronic structure of $(TMTSF)_2PF_6$ [53] show that anion potential has only a minor influence on the conduction band dispersion. The anions could also interact indirectly with the $\pi$ conduction holes via the polarization of the methyl groups and the $\sigma$ bond skeleton as outlined in the last section and previously discussed in references [51, 52, 13]. The importance of this last effect has not yet been evaluated. Another important indirect effect is the lattice deformation caused by the libration of the anions and methyl groups and their low-temperature H-bond linkage. In particular we have shown that the anion libration is the main contributor to the lattice thermal expansion. The main effect of the anion's displacement and of its coupling to the TMTSF lattice degrees of freedom is to induce a shear deformation of the methyl group cavity. This will explicitly deform the ($a$, $b$) TMTSF layer by changing not only the intra- and inter-stack distances, but also the relative angular position between neighbouring TMTSF molecules. Such shear deformations have a drastic influence on both the amplitude and sign of the transfer integrals along the different conducting paths and finally on the warping of the Fermi surface and its density wave nesting properties [53].

Finally the accurate analysis reported here for the Bechgaard salts could be generalized to other organic conductors such as the $(BEDT-TTF)_2X$ salts which present also disordered lattice degrees of freedom which include both the anionic layers and the conformation of the ethylene terminal groups of the BEDT-TTF molecule. This is for example the case of the $T_c$ = 11.6 K ambient-pressure superconductor $\kappa$-$(BEDT-TTF)_2Cu[N(CN)_2]Br$ which exhibits a freezing of the ethylene group motion below 200 K [54] together with a mysterious glass-like transition at $T_g$ ~77 K [55]. Interestingly the electronic properties of this salt depend significantly upon the cooling rate in the vicinity of $T_g$ [3]. Although it has been shown [56] that the glass-like transition is not primarily caused by the configuration freezing-out of the ethylene groups, additional disorders inside the H-bond network and its linkage to the anionic layer could be at the origin of the metastable states achieving the glassy behaviour. In relationship with the present study it should be interesting to accurately measured the thermal variation of the Bragg reflection intensity as a function of the cooling rate in the vicinity of $T_g$ for deuterated κ-(BEDT-TTF)$_2$X salts. This allows appreciating how tiny modifications in the H-bond network could perturb the BEDT-TTF packing. In particular disorder in the H-bond network could texture the conducting layer into domains where BEDT-TTF is slightly differently packed and ordered. In this situation the formation of non-superconducting disorder-induced regions should induce a Josephson coupling between more ordered superconducting domains, which could be an essential ingredient to explain the unusual dependence of of superconducting properties of κ-(BEDT-TTF)$_2$X with the amount of disorder[57].

**Acknowlegements**

P. Auban-Senzier is thanked for the conductivity measurements, as well as C. Bourbonnais, E. Canadell and M. Dressel for useful discussions. M. de Souza acknowledges financial support from the São Paulo Research Foundation -- Fapesp (Grants No. 2011/22050-4) and National Counsel of Technological and Scientific Development -- CNPq (Grants No. 308977/2011-4).

|        | $(TMTSF)_2PF_6$-D12 | $(TMTSF)_2PF_6$-H12 |
|---|---|---|
| $a$ (Å) | 7.1304(9) | 7.077(2) |
| $b$ (Å) | 7.6660(8) | 7.632(1) |
| $c$ (Å) | 13.3735(13) | 13.322(2) |
| $\alpha$ (°) | 84.173(4) | 84.14(2) |
| $\beta$ (°) | 88.044(5) | 88.00(2) |
| $\gamma$ (°) | 70.013(4) | 70.15(2) |
| $V$(Å$^3$) | 683.44(13) | 673.25 |

Table 1: Comparison of the 4 K lattice parameters and unit cell angles of $(TMTSF)_2PF_6$-D12 and -H12. The $(TMTSF)_2PF_6$-H12 data are taken from [23].

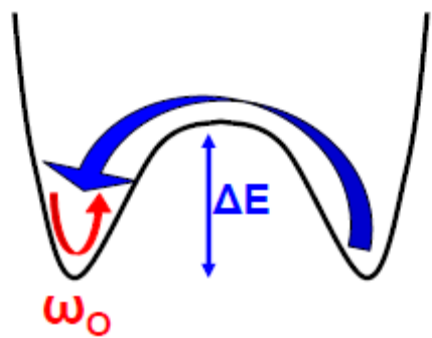

Figure 1. (Colour online) Schematic representation of the two types of molecular motion considered in the paper: thermally activated reorientation jumps across a potential barrier of height ΔE (in blue) and oscillations of frequency $\omega_0$ around the minimum of a potential well (in red).

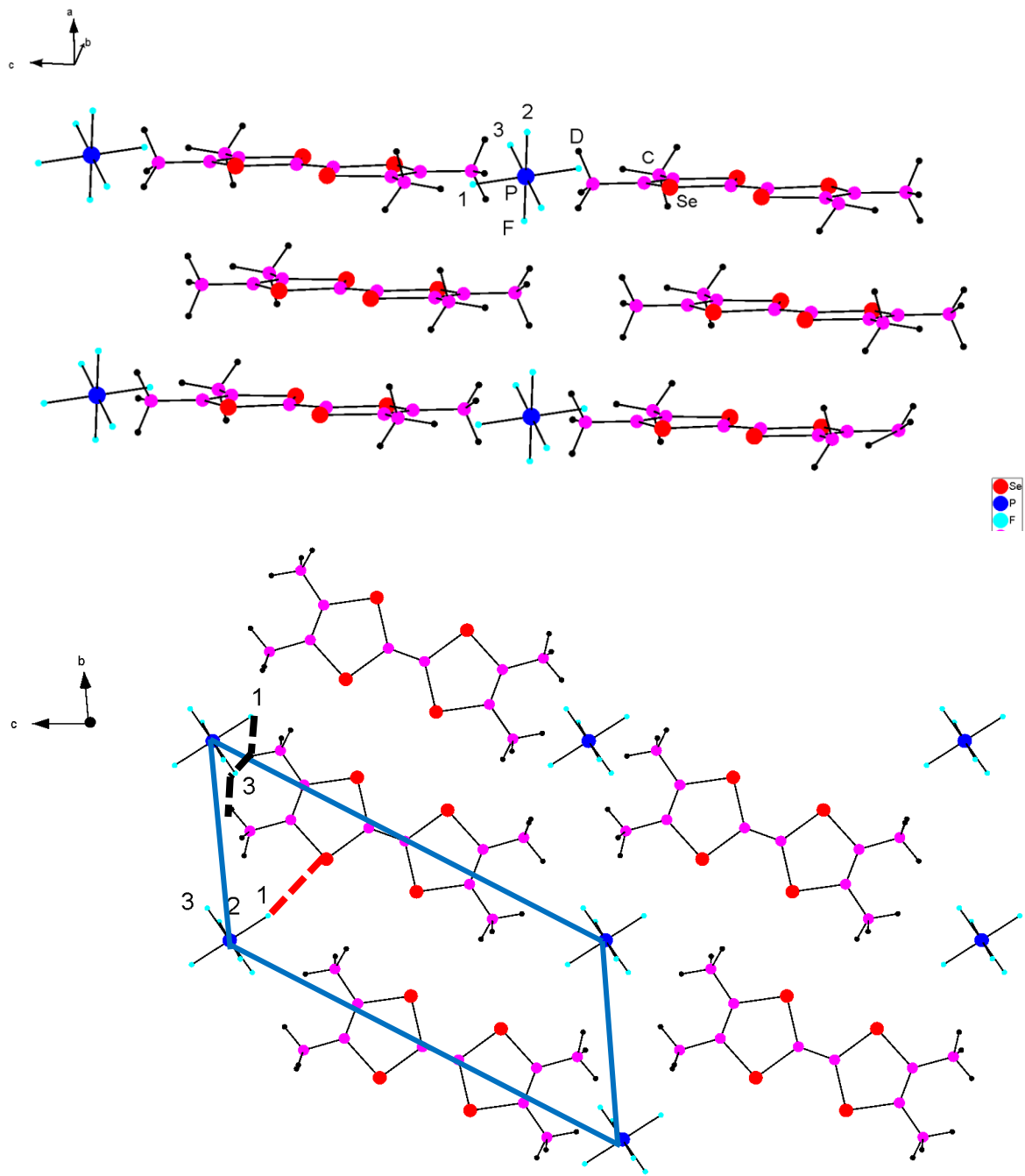

Figure 2. (Colour online) Structure of (TMTSF)$_2$PF$_6$-D12 at 4.2K. Top panel: perspective view from the ($a$, $c$) plane. Bottom panel: projection in the ($b$, $c$) plane whose unit cell is drawn. The F atoms are labelled. In the bottom panel the three short F-D distances and the short F-Se ($d_{12}$) distance are shown by the interrupted black and red lines, respectively.

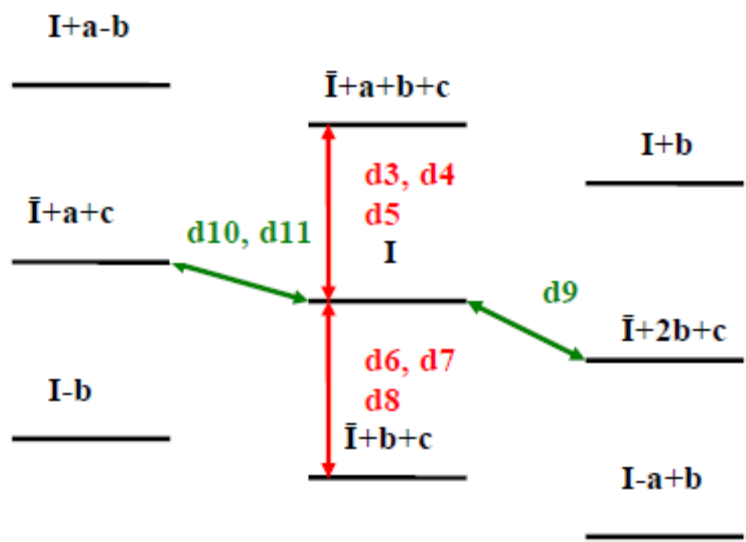

Figure 3: (Colour online) Schematic representation of the (*a*, *b*) layers of TMTSF together with the symmetry operations relating the molecules. The location of the shortest Se-Se distances given in table A.3 is indicated. The TMTSF molecule, whose atomic coordinates are given in table A.2, is labelled by I. Ī is related to I by inversion symmetry on the P site located at the origin of the unit cell. The stack is dimerized between molecules I and Ī+a+b+c.

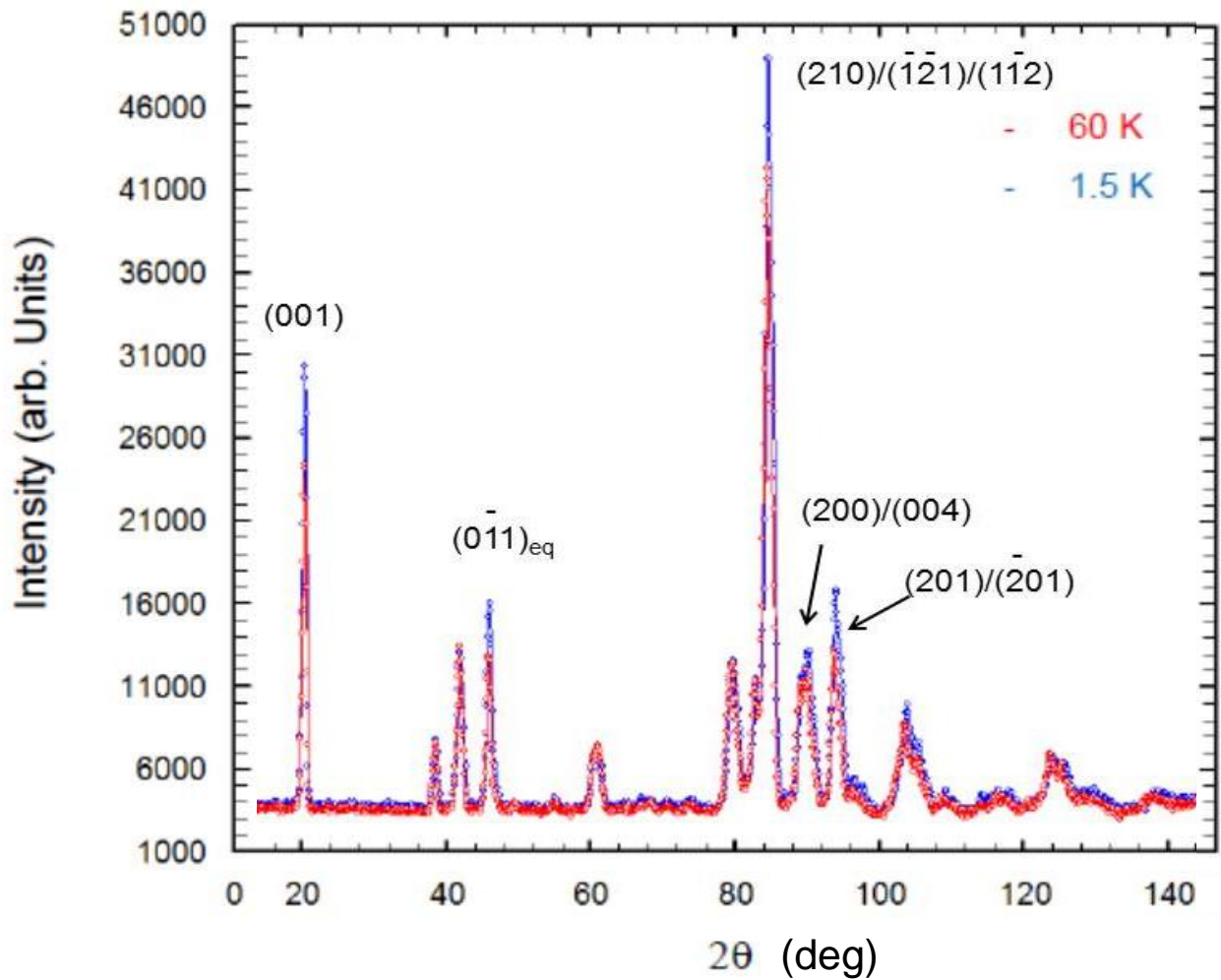

Figure 4. (Colour online) Neutron diffraction powder patterns ($\lambda = 4.741$ Å) of $(TMTSF)_2PF_6$-D12 at 1.5 K (blue points) and 60 K (red points). Some Bragg reflections experiencing the largest intensity variation are indexed.

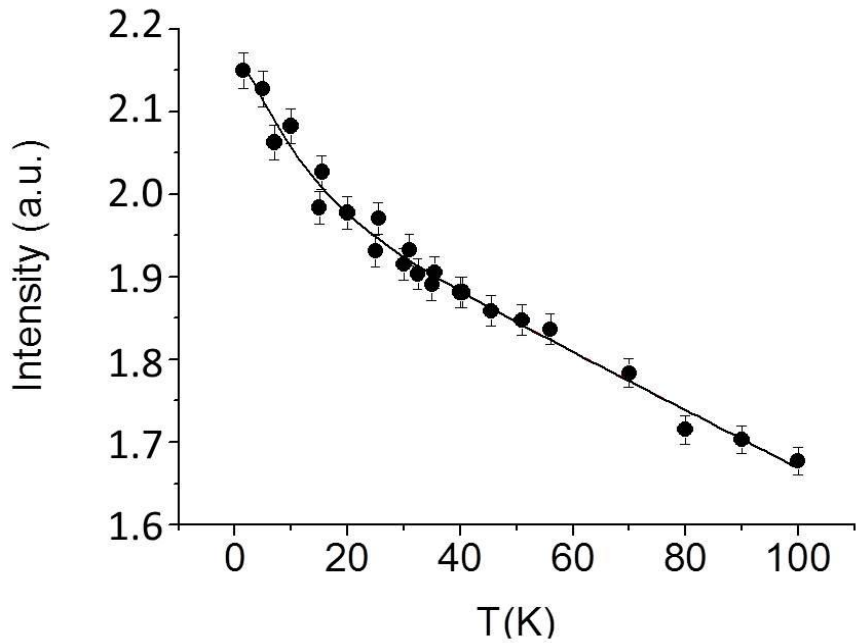

Figure 5 Thermal dependence of the integrated intensity of the (001) reflection of $(TMTSF)_2PF_6$-D12 (dots) and fit (continuous line) of the data by the expression (3) with $<\delta\varphi^2>_0 = 0.1$ rad² and $\theta_E = 8.3$ K.

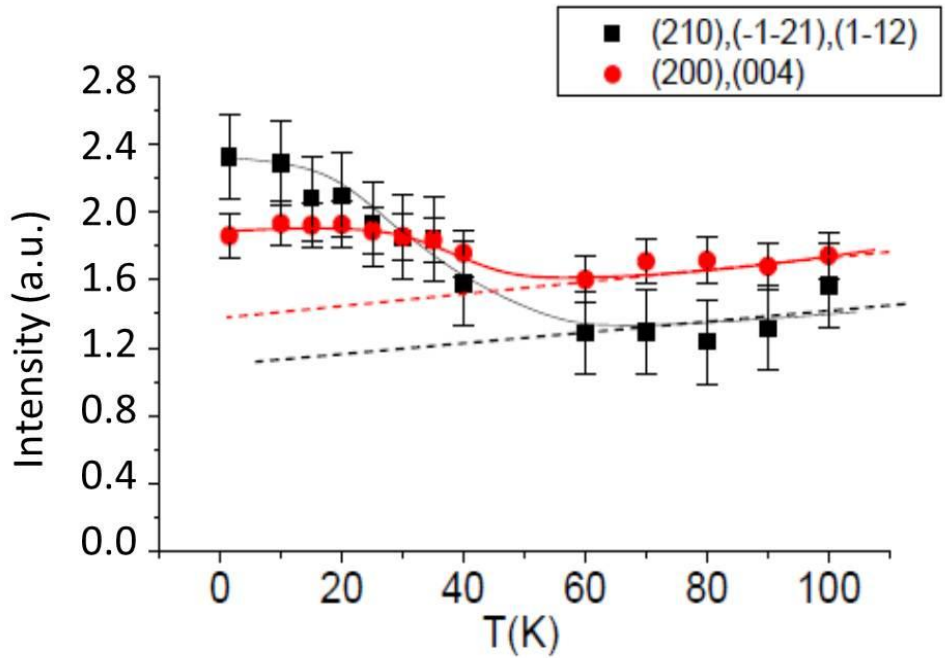

Figure 6. (Colour online) Thermal dependence of the integrated intensity of the two sets of {(210), (-1-21), (1612)} and {(200), (004)} reflections of $(TMTSF)_2PF_6$-D12 (black squares and red dots, respectively). The continuous line is a guide to the eyes. The dashed lines extrapolate the high-temperature dependence towards the low temperatures.

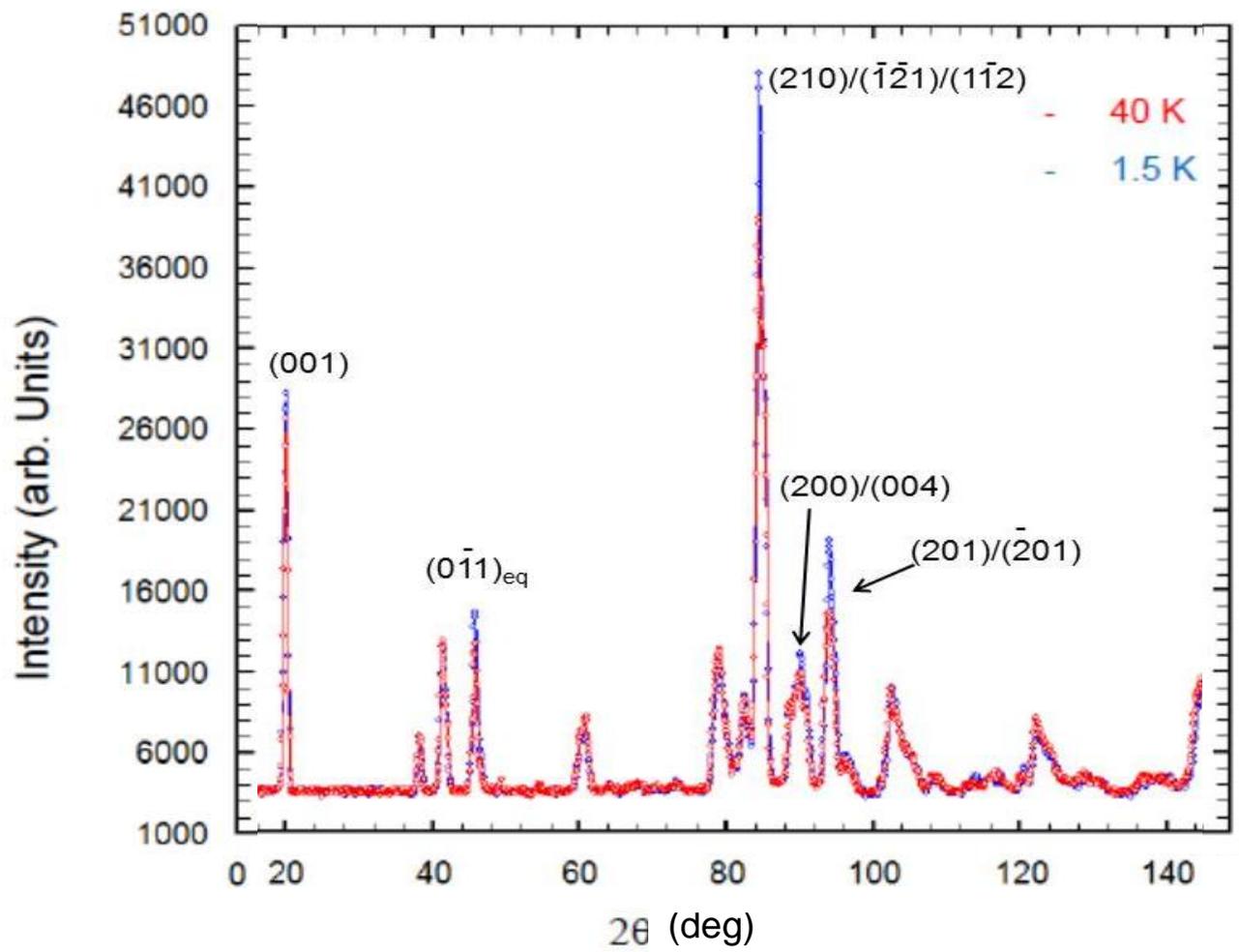

Figure 7. (Colour online) Neutron diffraction powder patterns ($\lambda = 4.741$ Å) of (TMTSF)$_2$AsF$_6$-D12 at 1.5 K (blue points) and 40 K (red points). Some Bragg reflections experiencing the largest intensity variation are indexed.

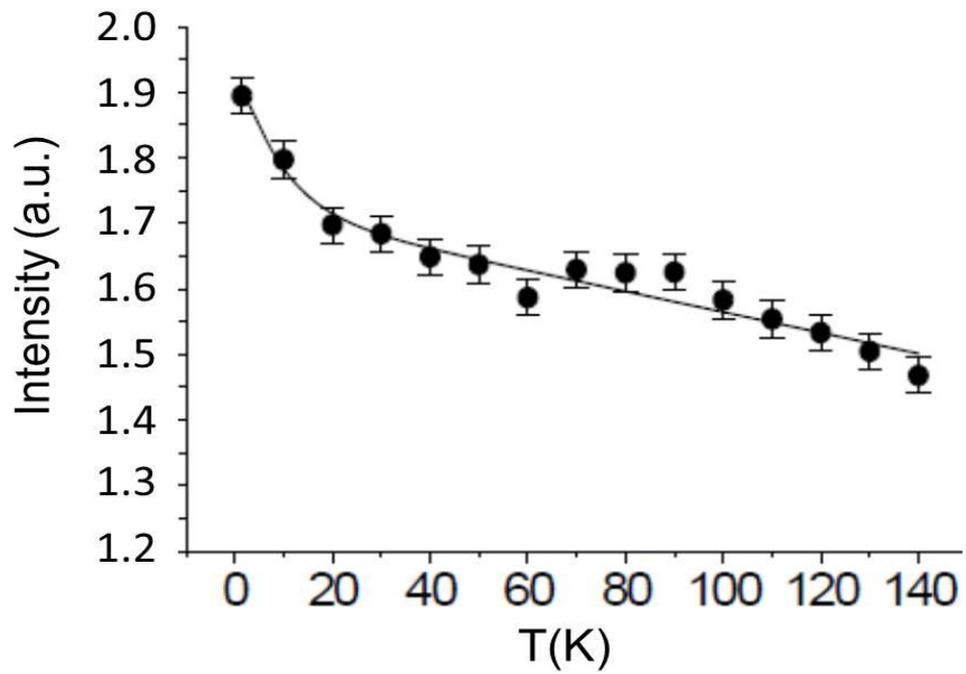

Figure 8. Thermal dependence of the integrated intensity of the (001) reflection of $(TMTSF)_2AsF_6$-D12 (dots) and fit (continuous line) of the data by expression (3) with $\langle\delta\varphi^2\rangle_0 = 0.1$ rad² and $\theta_E = 6.7$ K.

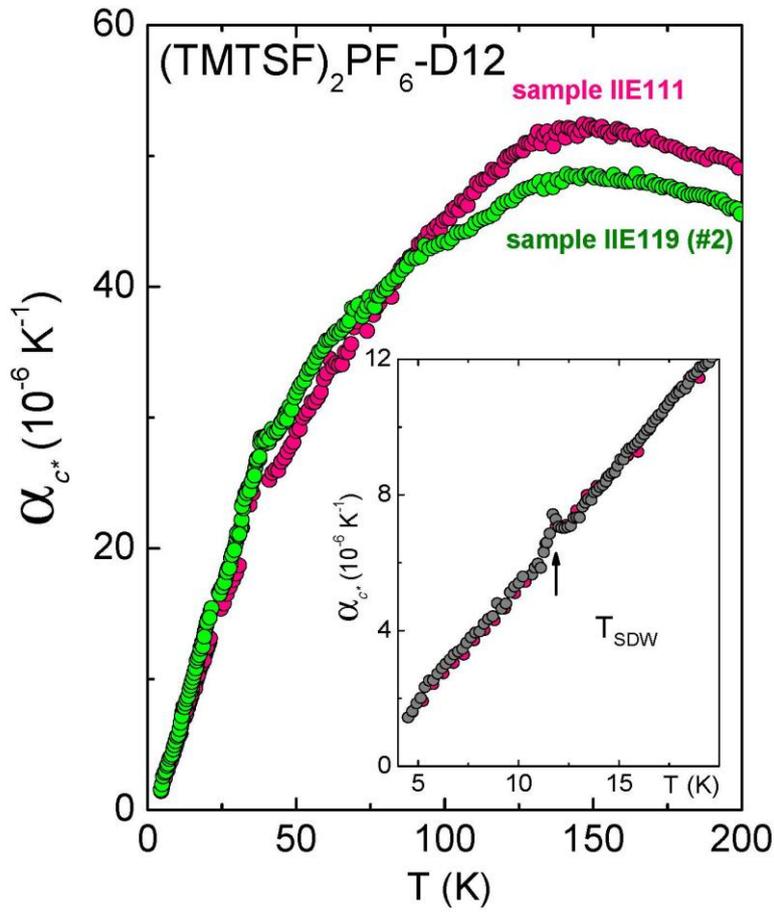

Figure 9. (Colour online) Uniaxial thermal expansion coefficient versus temperature measured along the interlayer direction $c^*$ for two $(TMTSF)_2PF_6$-D12 single crystals. The inset is a blow-up of the low-temperature data showing the thermal expansion anomaly at the SDW transition.

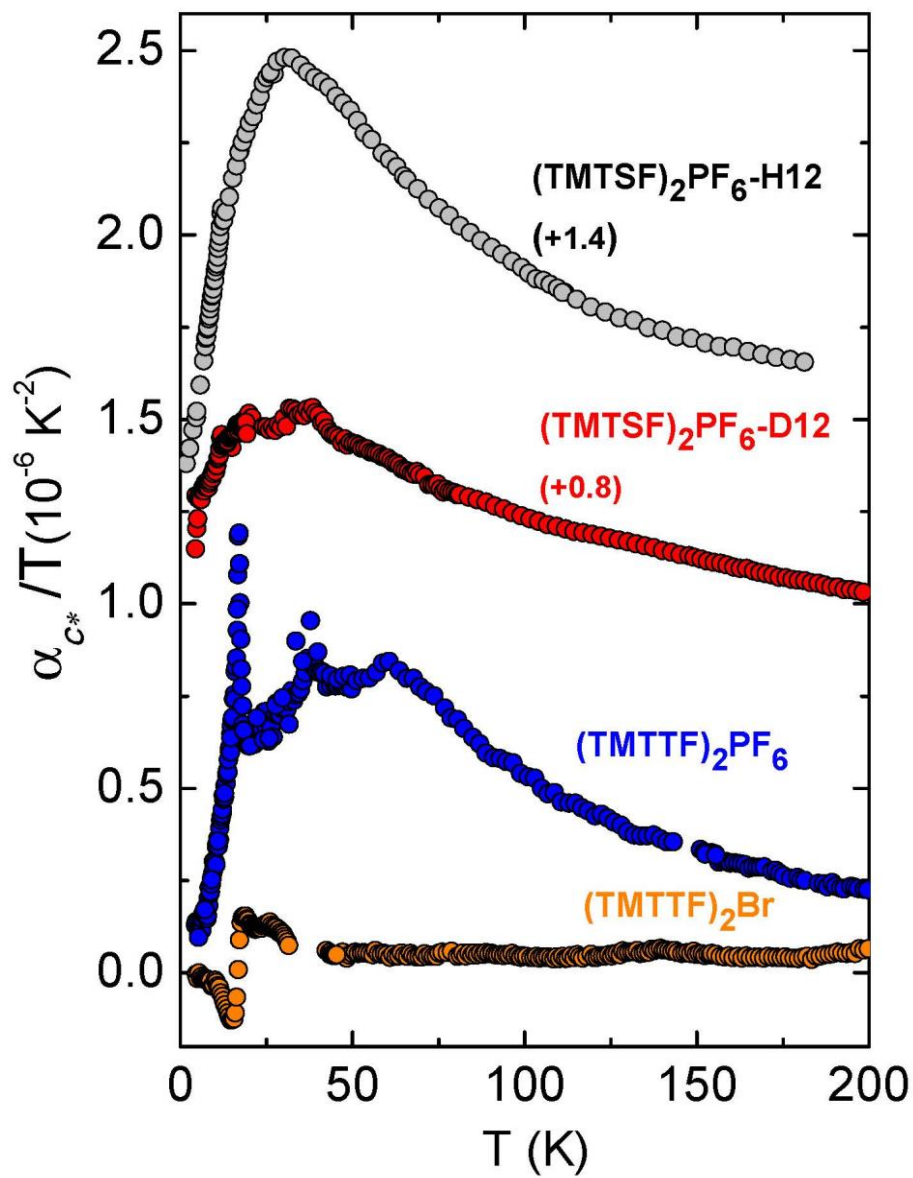

Figure 10. (Colour online) $\alpha_{c*}/T$ versus $T$ for (TMTSF)$_2$PF$_6$-H12, (TMTSF)$_2$PF$_6$-D12, (sample IIE119), (TMTTF)$_2$PF$_6$-H12 and (TMTTF)$_2$Br-H12. The data of the TMTSF salts are shifted for clarity by a constant (in units of $10^{-6}$ K$^{-2}$) indicated in the figure.

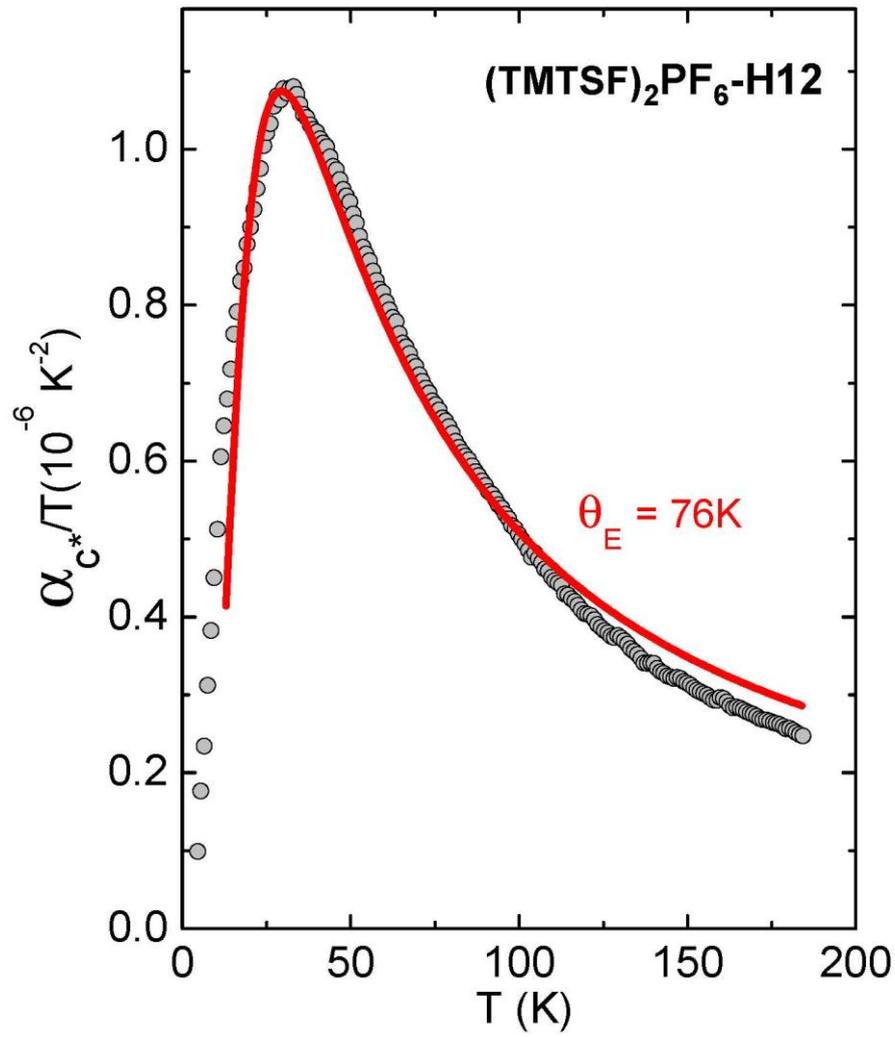

Figure 11. (Colour online) Thermal dependence of $\alpha_{c*}/T$ in $(TMTSF)_2PF_6$-H12 and its fit with one Einstein oscillator with $\theta_E = 76K$.

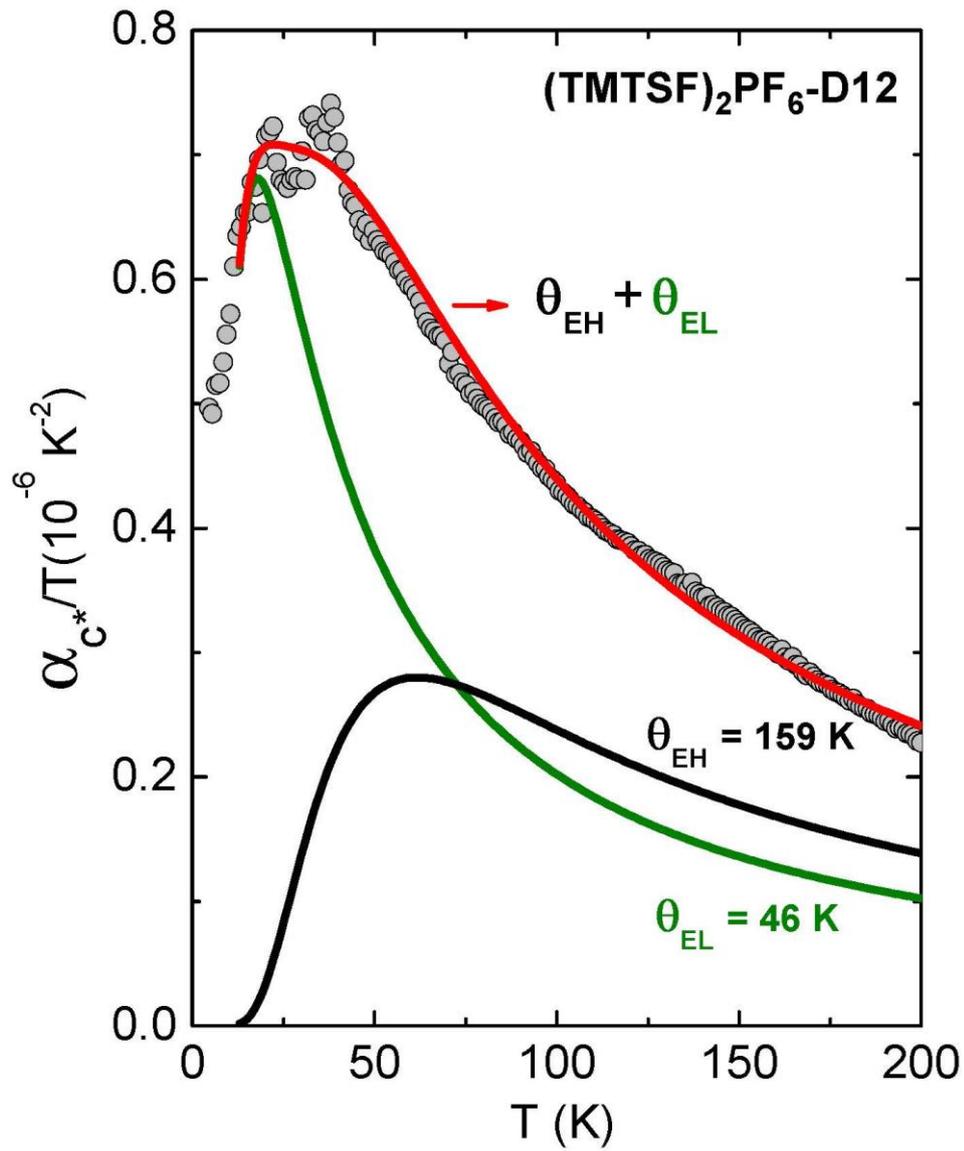

Figure 12. (Colour online) Thermal dependence of $\alpha_{c*}/T$ in $(TMTSF)_2PF_6$-D12 (sample IIE.119) and its fit with two Einstein oscillators (red curve). The green and black curves correspond to the contributions from the Einstein oscillators with low ($\theta_{EL}$) and high ($\theta_{EH}$) Einstein temperature, respectively.

# Annex A: Crystallographic data of (TMTSF)$_2$PF$_6$-D12 at 4.2K

| Empirical formula | Se$_8$C$_{20}$D$_{24}$. PF$_6$ |
|---|---|
| Temperature | 4.2K |
| Crystal system, Space group | Triclinic, P$\bar{1}$ |
| Unit cell dimensions | $a$ = 7.1304(9) Å    $\alpha$ = 84.173(4)°<br>$b$ = 7.6660(8) Å    $\beta$ = 88.044(5)°<br>$c$ = 13.3735(13) Å   $\gamma$ = 70.013(4)° |
| Volume | 683.44(13) Å$^3$ |
| Z, Calculated density | 1, 3.320 Mg/m$^3$ |
| Neutron wavelength | 0.840 Å |
| Theta range for data collection | 3.59 to 30.16° |
| Limiting indices | -3 ≤ h ≤ 7   -9 ≤ k ≤ 8   -15 ≤ l ≤ 15 |
| Reflections collected / unique | 2429 / 1891  [R(int) = 0.0179] |
| Completeness to theta = 30.16° | 77.2% |
| Refinement method | Full-matrix least-squares on F² |
| Data / restraints / parameters | 1891 / 0 / 269 |
| Goodness of fit on F² | 1.101 |
| Final R indices [I>2σ(I)] | R1 = 0.0481, wR2 = 0.1077 |
| R indices (all data) | R1 = 0.0537, wR2 = 0.1101 |
| Extinction coefficient | 3.43(15) |

Table A1. Crystal data and parameters of the structure refinement.

| Atom | x | y | z | U(eq) (Å²) |
|---|---|---|---|---|
| Se1  | 0.3065(3)  | 0.3354(2)  | 0.6160(1)   | 0.002(1) |
| Se2  | 0.1609(3)  | 0.7508(2)  | 0.5114(1)   | 0.001(1) |
| Se11 | 0.3755(3)  | 0.1687(3)  | 0.3847(1)   | 0.002(1) |
| Se12 | 0.2217(3)  | 0.5847(2)  | 0.28046(1)  | 0.001(1) |
| P    | 0.00000    | 0.00000    | 0.00000     | 0.002(1) |
| F1   | 0.0167(4)  | -0.1229(4) | 0.1073(2)   | 0.005(1) |
| F2   | 0.2362(4)  | -0.1023(4) | -0.0183(2)  | 0.006(1) |
| F3   | 0.0480(4)  | 0.1597(4)  | 0.0537(2)   | 0.006(1) |
| C1   | 0.2318(3)  | 0.5356(3)  | 0.6997(2)   | 0.003(1) |
| C2   | 0.1700(3)  | 0.7127(3)  | 0.6545(2)   | 0.003(1) |
| C3   | 0.2524(3)  | 0.4930(3)  | 0.4942(2)   | 0.003(1) |
| C4   | 0.1040(4)  | 0.8891(3)  | 0.7060(2)   | 0.003(1) |
| C5   | 0.2473(4)  | 0.4762(3)  | 0.8110(2)   | 0.003(1) |
| C11  | 0.3736(3)  | 0.2068(3)  | 0.2413(2)   | 0.003(1) |
| C12  | 0.3077(3)  | 0.3842(3)  | 0.1969(2)   | 0.003(1) |
| C13  | 0.2804(3)  | 0.4260(3)  | 0.4018(2)   | 0.003(1) |
| C14  | 0.2945(4)  | 0.4411(3)  | 0.0854(2)   | 0.004(1) |
| C15  | 0.4471(4)  | 0.0313(3)  | 0.1881(2)   | 0.003(1) |
| D101 | -0.0572(9) | 0.9586(8)  | 0.6983(4)   | 0.060(1) |
| D102 | 0.6093(6)  | -0.0371(6) | 0.1975(3)   | 0.039(1) |
| D103 | 0.1961(5)  | 0.5956(4)  | 0.8551(2)   | 0.023(1) |
| D111 | 0.1585(6)  | 0.3861(5)  | 0.8320(3)   | 0.033(1) |
| D112 | 0.4024(7) | 0.3959(6) | 0.8319(3) | 0.044(1) |
| D113 | 0.1376(5) | 0.8624(4) | 0.7862(2) | 0.025(1) |
| D121 | 0.1734(5) | 0.9880(4) | 0.6726(3) | 0.026(1) |
| D122 | 0.1419(9) | 0.480(1) | 0.0592(4) | 0.067(2) |
| D123 | 0.3377(6) | 0.5627(5) | 0.0655(2) | 0.027(1) |
| D131 | 0.3876(6) | 0.3286(5) | 0.0439(2) | 0.033(1) |
| D132 | 0.4152(5) | 0.0600(4) | 0.1081(2) | 0.023(1) |
| D133 | 0.3766(6) | -0.0688(5) | 0.2182(3) | 0.038(1) |

Table A2. Atomic coordinates and equivalent isotropic displacement parameters (x, y and z are expressed in fraction of *a*, *b* and *c* lattice parameters defined in table 1). The labelling of the atoms is the same as in ref. [23]. In this table U(eq), defined as one third of the trace of the orthogonalized $U_{ij}$ tensor, is related to the Debye-Waller factor of atom (nucleus) n of the Bragg reflection located at **G** by the expression $W_n = G^2 U_n(eq)/2$.

| | D12 at 4.2 K | H12 at 4 K | H12 at 1.7 K and $7 \times 10^2$ MPa |
|---|---|---|---|
| Se(1)-Se(12): $d_3$ | 3.938(7) | 3.915(7) | 3.872(4) |
| Se(2)-Se(11): $d_4$ | 3.888(7) | 3.861(7) | 3.815(4) |
| Se(1)-Se(2): $d_5$ | 3.978(9) | 3.943(7) | 3.893(4) |
| Se(1)-Se(12): $d_6$ | 3.836(7) | 3.799(7)? | 3.765(4) |
| Se(2)-Se(11): $d_7$ | 3.892(7) | 3.866(7)? | 3.828(4) |
| Se(1)-Se(2): $d_8$ | 4.071(9) | 4.054(7) | 4.004(4) |
| Se(2)-Se(2): $d_9$ | 3.712(7) | 3.711(7) | 3.669(4) |
| Se(1)-Se(11): $d_{10}$ | 3.754(7) | 3.737(7) | 3.695(4) |
| Se(11)-Se(11): $d_{11}$ | 3.870(7) | 3.857(7) | 3.826(4) |
| Se(12)-F(1): $d_{12}$ | 3.076(9) | 3.060(9) | 3.036(6) |

Table A.3: Se-F, inter-stack and intra-stack Se-Se distances ($d_i$ according to their label in figure 3) in $(TMTSF)_2PF_6$-D12 (this work), $(TMTSF)_2PF_6$-H12 (from ref. [23]) and in pressurized $(TMTSF)_2PF_6$-H12 (from ref. [37]). The labelling of Se atoms is given in table A.2. All distances are in Å. The distances $d_6$ and $d_7$ given in ref. [23] have been interchanged.

It is useful to compare the inter-molecular distances for the PF$_6$-D12 and PF$_6$-H12 structures. Table A.3 gives the shortest Se-Se inter-molecular distances defined in figure 3 for the 4 K and ambient-pressure structures of PF$_6$-D12 and PF$_6$-H12 as well as for the 1.7 K structure of PF$_6$-H12 taken at a pressure of $7\times10^2$ MPa.

The intra- and inter-dimer distances defined in figure 3, ($d_3$, $d_4$, $d_5$) and ($d_6$, $d_7$, $d_8$) respectively, expand on going from PF$_6$-H12 to PF$_6$-D12, as expected from the application of a negative pressure to PF$_6$-H12. The inter-stack distance $d_9$ does not change between PF$_6$-H12 and PF$_6$-D12 while the inter-stack distances ($d_{10}$, $d_{11}$) expand less than expected from varying the pressure to the PF$_6$-H12 salt. This last feature is expected regarding the relative variation of the *b* parameter which is twice smaller than the *a* parameter upon deuteration (table 1). The shortest Se(12) – F(1) contact distance $d_{12}$ (shown in the bottom panel of figure 2) also increases upon deuteration as expected from the application of a negative pressure.

## Annex B: Rotational disordering of a methyl group

We consider a simplified system made of a single D atom (which mimics a methyl group – see below) attached to a molecule of structure factor $F_{core}(\mathbf{Q})$. With the distances defined in figure B.1, the structure factor of this entity is:

$$F(\mathbf{Q})= F_{core}(\mathbf{Q})+ b_D \exp(i\mathbf{QR}) = F_{core}(\mathbf{Q})+ b_D \exp(i\mathbf{Qd}) \exp(i\mathbf{Qr}) \,, \qquad (B.1)$$

where $b_D$ is the scattering length of the D located in $\mathbf{R}= \mathbf{d}+\mathbf{r}$. Within the orthonormal (x, y, z) referential, shown in figure B1, $\mathbf{d}$ is along the z axis of the cone containing the methyl group. D is located in the circle of radius r located in the (x, y) plane; $\varphi$ is the angle between $\mathbf{r}$ and the fixed direction y. With these coordinates the D disorder is described by the variable $\varphi$.

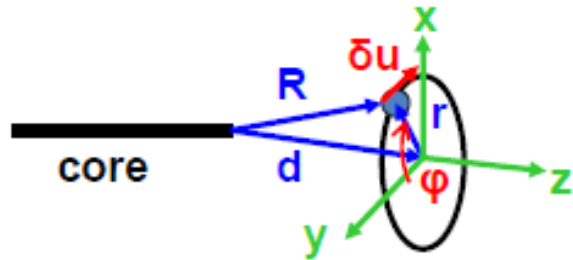

Figure B1. Coordinates of the disordered D atom

Using the reciprocal wave vector $\mathbf{Q'}=h\mathbf{a'}*+l\mathbf{c'}*$ expressed in $\mathbf{a'}*$ and $\mathbf{c'}*$ units related to the y and z directions, expression (B.1) becomes for
  - wave vectors along the cone axis $\mathbf{Q'}=l\mathbf{c'}*$:
    $$F(l)= F_{core}(l)+ b_D \exp(2i\pi l d/c'), \qquad (B.2)$$
  - wave vectors perpendicular to the cone axis $\mathbf{Q'}=h\mathbf{a'}*$:
    $$F(h)= F_{core}(h)+ b_D \exp(2i\pi h(r\sin\varphi)/a'). \qquad (B.3)$$

The disorder leads to a distribution of φ. If P(φ) is the normalized distribution of φ, (B.3) becomes:

$$F(h) = F_{core}(h) + b_D \int \exp(2i\pi h(r\sin\varphi)/a') P(\varphi) d\varphi. \quad (B.4)$$

For a total disorder where all the φ values are equi-probable the integral in the second member of (B.4) vanishes and:

$$F(h) = F_{core}(h). \quad (B.5)$$

The D does not contribute to the structure factor. For D ordered in $\varphi_0$ (B.4) becomes:

$$F(h) = F_{core}(h) + b_D \exp(2i\pi h(r\sin\varphi_0)/a'). \quad (B.6)$$

The D contributes to the structure factor (B.4). If P(φ) changes in temperature (B.4) will be thermally dependant. Thus when the D disorder increases upon heating the structure factor will progressively vary from (B.6) to (B.5).

Note that in our simplified model the disorder does not modify the projection of D on the cone axis: F(l) is independent of φ, and thus the structure factor (B.2) does not change in temperature.

This very simple model allows calculating more quantitatively the effect of a partial disorder. Using the relationship:

$$\exp(i u\sin\varphi) = \sum_n J_n(u) \exp(in\varphi),$$

where $J_n$ are the Bessel functions with n being a positive or negative integer, (B.4) becomes:

$$F(h) = F_{core}(h) + b_D \sum_n J_n(2\pi hr/a') \int P(\varphi) \exp(in\varphi) d\varphi,$$

$$F(h) = F_{core}(h) + b_D \sum_n J_n(2\pi hr/a') P(n), \quad (B.7)$$

where $P(n) = \int \exp(in\varphi) P(\varphi) d\varphi$ is the Fourier transform of P(φ).
If P(φ) is a Gaussian distribution centred on $\varphi_0$ and of variance $\sigma = \sqrt{\langle(\varphi-\varphi_0)^2\rangle}$, P(n) is also a Gaussian:

$$P(n) = \exp(in\varphi_0) \exp(-n^2\sigma^2) \quad (B.8)$$

Far from the complete order the dominant contribution will come from the n=±1 and n=±2 terms which give respectively imaginary and real components at (B.7) which now becomes:

$$F(h) = F_{core}(h) + 2ib_D\sin(\varphi_0)J_1(2\pi hr/a')\exp(-\sigma^2) + 2b_D\cos(2\varphi_0)J_2(2\pi hr/a')\exp(-4\sigma^2) \quad (B.9)$$

If now one considers explicitly a methyl group which rotates rigidly, $b_D$ must be replaced by the methyl group structure factor $F_{D3}$. However as the three D of the methyl group are located at angles φ, φ+2π/3 and φ-2π/3 which enters in the exp(inφ) term of (B.7), the methyl group structure factor $F_{D3}$ splits into n components $F_{D3}(n)$. Then the expression analogous to (B.7) becomes:

$$F(h) = F_{core}(h) + \sum_n F_{D3}(n) J_n(2\pi hr/a') P(n), \quad (B.10)$$

where:

$$F_{D3}(n) = b_D (1 + 2\cos(2n\pi/3)).$$

This does not change the general form of the expression (B.9).

In $(TMTSF)_2 X$ there are four methyl groups per TMTSF, located on two types of cones differently oriented, and two TMTSF molecules per unit cell related by inversion symmetry. As there are two different cone orientations, P(n) should contribute to the structure factor F(**Q**) whatever the **Q** orientation. Thus for a general wave vector **Q**, F(**Q**) will depend upon P(n) as in the expression (B.10). If P(φ) and thus P(n) vary in temperature, the intensity of the Bragg reflection located in **G** and given by:

$$I(\mathbf{G}) = |F(\mathbf{Q})|^2 \delta(\mathbf{G}-\mathbf{Q}), \tag{B.11}$$

will be thermally dependant.

The contribution of the disorder of all the methyl groups of the unit cell at the intensity of the Bragg reflections should keep the inversion symmetry of the average structure. Thus the imaginary contribution of individual methyl group disorder, present in (B.9), should cancel because the total structure factor F(**G**) entering in the Bragg intensity (B.11) must be real. If σ is large (this is the case when the temperature increases) the leading contribution of the methyl group disorder should be proportional to $\exp-4\sigma^2$. This contribution, which adds to the core structure factor, corrects the Bragg intensity I(**G**) by δI(**G**) proportional to $\exp-4\sigma^2$ (expression (1) in the main text). In this simple model the thermal dependence of σ in the exponential term describe the thermal variation of the Bragg intensity.

## Annex C: Thermodynamics of Einstein oscillators

Let consider a collection of N independent Einstein harmonic oscillators of quantized energy $E_n = (n+1/2)\hbar\omega_0$. If one poses $k_B\theta_E = \hbar\omega_0$, the partition function per oscillator is:

$$Z = [2\text{sh}(\theta_E/2T)]^{-1}. \tag{C.1}$$

The free energy per oscillator, F, is:

$$F = -k_B T \ln z = k_B T \ln[2\text{sh}(\theta_E/2T)]. \tag{C.2}$$

The energy per oscillator is:

$$u = (k_B T^2)\partial \ln z/\partial T = \hbar\omega_E \langle x^2 \rangle_T = \hbar\omega_0 \coth(\theta_E/2T). \tag{C.3}$$

(C.3) defines the mean square fluctuation of the oscillator variable x:

$$\langle x^2 \rangle_T = \langle x^2 \rangle_0 \coth(\theta_E/2T). \tag{C.4}$$

The entropy per oscillator is given by:

$$s = -\partial f/\partial T = k_B \ln z + u/T. \tag{C.5}$$

The specific heat per oscillator is:

$$c_V = \partial u/\partial T = T\partial s/\partial T = k_B(\theta_E/2T)^2 [\text{sh}(\theta_E/2T)]^{-2}. \tag{C.6}$$

The successive thermal derivatives of the entropy are:

$$\partial s/\partial T = c_V/T = (2k_B/\theta_E)(\theta_E/2T)^3 [\text{sh}(\theta_E/2T)]^{-2}, \tag{C.7}$$

and:

$$\partial^2 s/\partial^2 T = T^{-1}(\partial c_V/\partial T - c_V/T) = (c_V/T^2)(\partial \ln c_V/\partial \ln T - 1). \tag{C.8}$$

If T << $\theta_E$/2: $\partial s/\partial T \approx k_B/T$, decreases with T, while if T >> $\theta_E$/2: $\partial s/\partial T \approx (k_B/\theta_E)(\theta_E/T)^3 \exp-(\theta_E/T)$, increases with T. Thus, between these two regimes $\partial s/\partial T$ exhibits a maximum at $T_M$ given by:

$$\partial^2 s/\partial^2 T = 0 \text{ or } \partial \ln c_V/\partial \ln T = 1.$$

$T_M$, solution of:

$$\text{th}(\theta_E/2T) = \theta_E/3T, \tag{C.9}$$

amounts to $T_M \approx 0.39\theta_E$, which is close to $\theta_E/2$.

# Annex D: Thermodynamics of the lattice expansion coefficient

The volumetric thermal expansion coefficient of a solid $\beta(T)$ is related to the isothermal compressibility $\kappa_T$ by the Grűneisen relationship:

$$\beta(T) = \kappa_T \, C_V(T) \, \Gamma/V, \qquad (D.1)$$

where $C_V(T)$ is the specific heat at constant volume (V) and

$$\Gamma = -(\partial \ln T/\partial \ln V)_S \qquad (D.2)$$

is the Grűneisen parameter. Using the thermodynamic relation (C7):

$$\partial S/\partial T = C_V(T)/T, \qquad (D.3)$$

one thus deduces that:

$$\beta(T)/T = (\partial S/\partial T)\, \kappa_T \, \Gamma/V \qquad (D.4)$$

is a measure of the thermal derivative of the entropy.

If one considers the unidirectional expansion coefficient $\alpha(T)$, the volume must be replaced by the sample length along which the measurement is performed. In this paper, we consider measurements along the interlayer direction $c^*$. In (TMTTF)$_2$PF$_6$ the thermal dependences of $\beta(T)/T$ and $\alpha_{c^*}(T)/T$ behave similarly in temperature [39]. This means that the thermal measurement along $c^*$ behaves as the volume dependence.

In our analysis of the data we consider only the lattice contribution to the thermal expansion. This contribution involves the lattice or phononic Grűneisen parameter $\Gamma_{ph}$ which takes into account the volume dependence of the vibration modes. $\Gamma_{ph}$ is generally taken as the sum of the volume dependence of independent vibration modes of frequency $\omega_i$. This allows defining the mode Grűneisen parameter:

$$\Gamma_{ph,i} = -\partial \ln \omega_i/\partial \ln V . \qquad (D.5)$$

The thermal expansion will thus be dominated by the phonon modes i having the highest $\Gamma_{ph,i}$. We show in the main text that this is the case for the libration modes of the anion in the Bechgaard and Fabre salts.